\documentclass[a4paper,11pt]{article}
\pdfoutput=1
\usepackage{graphicx}
\usepackage{jcappub}  
\usepackage[T1]{fontenc}
\usepackage{longtable}
\usepackage{float}
\usepackage{dcolumn}
\usepackage{bm}
\usepackage{appendix}
\usepackage{multirow}
\usepackage{color}
\usepackage{soul}
\usepackage[normalem]{ulem}
\usepackage[multiple]{footmisc}
\usepackage[utf8]{inputenc}
\usepackage{footmisc}
\usepackage{amsmath}

\title{Connecting early and late epochs by $f(z)$CDM cosmography}
\author[a,b]{Micol Benetti, } \emailAdd{benettim@na.infn.it}
\author[a,b,c,d]{Salvatore Capozziello} \emailAdd{capozzie@na.infn.it}
\affiliation[a]{Dipartimento di Fisica  ``E. Pancini", Universit\`a di Napoli  ``Federico II", Via Cinthia, I-80126, Napoli, Italy}
\affiliation[b]{Istituto Nazionale di Fisica Nucleare (INFN), sez. di Napoli, Via Cinthia 9, I-80126 Napoli, Italy}
\affiliation[c]{Gran Sasso Science Institute, Via F. Crispi 7, I-67100, L' Aquila, Italy}
\affiliation[d]{Laboratory for Theoretical Cosmology, Tomsk State University of Control Systems and Radioelectronics (TUSUR), 634050 Tomsk, Russia.}
\abstract{
The cosmographic approach is gaining considerable interest as a model-independent technique able to describe the late expansion of the universe. Indeed, given only the observational assumption of the cosmological principle, it allows to study the today observed accelerated  evolution of the Hubble flow without  assuming  specific cosmological models. 
In general,  cosmography is used to reconstruct the Hubble parameter as a function of the redshift, assuming an arbitrary fiducial value for the current matter density, $\Omega_m$,  and analysing low redshift cosmological data.
Here we propose a different strategy, linking together the parametric cosmographic behavior of the late universe expansion with the small scale universe.
In this way, we do not need to assume any ``a priori" values for the cosmological parameters, since these are constrained at early epochs using both the  Cosmic Microwave Background Radiation (CMBR) and Baryonic Acoustic Oscillation (BAO) data. In other words, we want to develop a cosmographic approach without assuming any background model but considering a $f(z)$CDM model where the function $f(z)$ is given by a suitable combination of polynomials capable of tracking the cosmic luminosity distance, replacing the cosmological constant $\Lambda$.
In order to  test  this  strategy, we  describe the late expansion of the universe using the Pad\'e polynomials. Specifically, we adopt a $P(2,2)$ series, that is a promising rational series which guarantees a good convergence also at high redshift. This approach is discussed in the light of the recent $H(z)$ values indicators, combined with Supernovae Pantheon sample, galaxy clustering and early universe data, as CMBR and BAO.
We found an interesting dependence of the current matter density value with cosmographic parameters, proving the inaccuracy of setting the value of $\Omega_m$ in cosmographic analyses. 
Furthermore, a non-negligible effect of the cosmographic parameters on the CMBR temperature anisotropy power spectrum is shown, and constraints by selected joint datasets are reported.
Finally, we found that the cosmographic series, truncated at third order, shows a better $\chi^2$ best  fit value then the {\it vanilla} $\Lambda$CDM model. This can be interpreted as the  requirement that higher order corrections have to be considered to correctly describe low redshift data and remove the degeneration of the models. 
}

\begin{document}
\maketitle

\section{Introduction}
 
The mechanism driving the accelerated expansion of the Universe at early  and  late times represent one of  the most important open problems in cosmology and fundamental physics. Currently, the idea of a crisis of the flat $\Lambda$CDM  model is getting stronger and stronger as a result of the tensions emerged between the $H_0$ measurements derived by constraints with early epoch data~{\cite{Ade:2015xua, Aghanim:2018eyx} and direct measured using different geometric distance calibrations of Cepheids~\cite{Riess:2019cxk}. Also, others independent measurements, with no shared observational systematics, as DES Collaboration results \cite{Abbott:2017smn}, SPT Collaboration \cite{Henning:2017nuy},  H0LiCOW collaboration \cite{Bonvin:2016crt}, and others reinforce the idea of the existence of a tension that does not depend on experimental systematics but on the physics of the cosmological scenario. 

Let we stress that the standard cosmological model describes a homogeneous and isotropic universe composed of about $95\%$ of dark components, of which neither physics nor a fundamental description is known. Although it  passed several cosmological tests and it is able to describe several cosmological observations, actually we should not be too surprised if it shows some troubles in representing  the most recent universe, that is thought to be dominated by dark energy (or $\Lambda$) in order to explain the current universe expansion.
Several alternative theories to the dark energy  assumption have been developed, such as the extended theory of gravity, quintessence models or approaches with non-minimal couplings 
\cite{Capozziello:2002rd, Hu, Starobinsky:2007hu, Capozziello:2011et, Nojiri:2017ncd}. 
Finding the correct model for the late-time acceleration turned out to be not easy at all, so that  the theoretical efforts to find a dynamic model   describing  the current observations, i.e. $\Lambda$ given by a barotropic fluid, have been placed side by side to kinematic models where the cosmological constant is simply assumed to be a function of the cosmic time (or the scale factor $a(t)$) \cite{Sahni:1999gb}. Indeed, although kinematic and dynamical models of $\Lambda$ lie on completely different theoretical bases, they may be equivalent, if a  space-time background is assumed to be described  by a  Friedmann-Lema\^itre-Robertson-Walker metric.
Then, it is possible to tackle the problem by an inverse scattering approach. It can be  based on reconstructing the Hubble parameter evolution $H(z)$ as a function of redshift, where we do not assume any specific dark energy or modified gravity model, and only subsequently search for some fundamental physics description.
In this context,  cosmography can be  a powerful tool to break the degeneracy among cosmological models, and it is currently widely adopted to understand universe's kinematics at late epochs.  
The  idea is to parameterize all quantities of interest through a Taylor series expansion around here and now, providing, in principle, a way to directly match with observations.
Noteworthy, the  description of the observations strongly depends on the amount of information: i.e.  it  directly depends on  where  the cosmographic series is truncated.
Specifically,  the second order expansion parameter, the deceleration parameter $q_0$, infers whether the universe is accelerating or decelerating. Including the third order, we can consider the change of the universe dynamics, while,  considering also the fourth order, we can discriminate between evolving dark energy term or a cosmological constant behavior.
Constraining these parameters is a tough challenge, especially looking at orders higher than the third, due to both the accuracy of  available data but also to the many assumptions. 

If on the one hand there are efforts to limit the approximations by choosing series of rational polynomials ever more efficient instead of simpler series, on the other hand,  one wonders if it is possible to constrain the assumptions, i.e. the value of cosmological parameters (e.g. the value of  non-relativistic matter, $\Omega_m$), needed to correctly express the Hubble rate.
Therefore, it seems natural to investigate the connection between low-redshift cosmographic parameterization with high redshifts cosmological model, building  up self-consistent models  allowing  a proper description of early universe 
and adopts a model-independent technique able to  describe the current observations at low redshift \cite{sahni}. 
Let us stress that  a cosmographic parameterization at large scales replaces the strict assumption of dark energy in the standard model, not having the issue to say what is driving the current accelerated expansion. 

In this work, we propose a new strategy. Instead of assuming $\Lambda$CDM as  background and studying dark energy parameterizations to deal with tensions, we consider a global cosmographic approach that we call $f(z)$CDM. Here,  a function of the redshift gives the cosmographic behavior of luminosity distance according to the convergence properties of a given class of polynomials, in principle at any epoch. 
Specifically, here, we take into account as an example Pad\'e polynomials that  overcome convergence limits of   standard cosmography, based on Taylor expansion \cite{Baker66, Aviles:2014rma, Capozziello:2018jya}. In particular, Pad\'e polynomials allow  a good series convergence even at redshift higher than $z=1$. The final purpose is to connect early and late universe by a model-independent cosmographic approach. The procedure is quite general depending only on the convergence of the polynomials chosen in $f(z)$.

The paper is organized as follow. In Sec. \ref{Sec:DL} we introduce the cosmographic approach in its basic formulation, while, in Sec. \ref{Sec:Pade}, we describe the Pad\'e series approximation.
The   dataset and codes adopted  for our analysis are discussed in Sec. \ref{Sec:Analysis}. Results are shown in Sec. \ref{Sec:Results}, where we present the observational constraints on  cosmographic parameters. Finally, Sec. \ref{Sec:Conclusions} we draw conclusions, give a summary,  and outline perspectives of this  strategy.

\section{Luminosity Distance and Cosmography}
\label{Sec:DL}
Let us start recalling the basic definitions of cosmography. 
The luminosity distance, $D_L$, can be written in terms of flux - luminosity relations or, 
in a flat universe, in terms of the redshift, $z$, of the observed object as

 \begin{eqnarray}
 \label{eq:DL_1}
D_L &=& (1+z) D(z)
    = (1+z) \frac{c}{H_0} \int^z_0 \frac{dz'}{H(z')},
 \end{eqnarray}
 with $D(z)$ the comoving distance depending  on the present value of the Hubble constant, $H_0$, and its evolution as a function of the redshift. 
The above formula can be simplified for low redshift values using the comoving distance relation by the Hubble - Leima\^itre law, where the recession velocity, $v$, is given by $v(z) = H_0 D(z)$. 
At the same time, it is worth noticing  that the simple definition of redshift from the Doppler formula $z \sim \frac{v(z)}{c}$ is valid for low velocity (i.e. low redshift), while for higher velocity we need to use the scale factor $a(t)$ as

\begin{equation}
z = \frac{a(t_0)}{a(t_e)} -1 , 
\label{eq:z_a}
\end{equation}
where $a(t_e)$ refers to the observed source (e.g. a galaxy) and  $a(t_0)=1$ is the normalized present value. 
Then, for low redshifts, we can write the luminosity distance relation of Eq.(\ref{eq:DL_1}) as

\begin{equation}
\label{eq:DL_lowz}
D_L = (1+z) \frac{z c}{H_0} .
\end{equation}
In order to consider higher $z$, we need to assume a cosmological model to make explicit the evolution of $H(z)$ in  Eq.(\ref{eq:DL_1}) in terms of universe fluids density and dynamic. 

Alternatively, we can express the luminosity distance in a model-independent parametric way by adopting the cosmographic approach, introduced as a  technique capable, in principle,  of tracing back the universe kinematics without the need of assuming  specific cosmological models.

{Let we stress that assuming a cosmological model to describe cosmic history is often biased by the fact that some constraints are assumed a priori, affecting the cosmic reconstruction and leaving the results not consistent with observations. On the other hand,  a model-independent reconstruction of the cosmic history should be based on a robust analysis of cosmological observations. In other words, data should be  able to provide reliable constraints on the behavior of cosmological  parameters, running with redshift, independently of any cosmological model or underlying gravity theory. 
For example, it has been shown that a principal component analysis,  adopting SNeIa data, can help to reconstruct the Hubble parameter \cite{emille}. This kind of  procedure  allows  to determine $H_0$ and, in principle, any other cosmographic parameter with reasonable uncertainty and  without  any ad-hoc parameterizations.
Also, inference algorithms can be adopted to reconstruct the cosmic expansion history using the information field theory, a statistical field theory  suited for the construction of optimal signal recovery algorithms  \cite{porqueres}. In both cases, the cosmic history is not assumed a priori but it is "reconstructed" by the optimization of data. In this sense,  cosmography  adopts kinematics to reconstruct cosmological dynamics. It can be considered a sort of "inverse scattering approach".}

Specifically,  we can expand the scale factor, that is the only degree of freedom governing the universe according to the cosmological principle, around the present epoch, that is 

\begin{equation}
a(t)=1+\sum_{k=1}^{\infty}\dfrac{1}{k!}\dfrac{d^k a}{dt^k}\bigg | _{t=t_0}(t-t_0)^k\ ,
\label{eq:scale factor}
\end{equation}
and define the \textit{Hubble}, \textit{deceleration}, \textit{jerk} and \textit{snap} parameters respectively as 

\begin{subequations}
\begin{align}
&H(t)\equiv \dfrac{1}{a}\dfrac{da}{dt} \ , \hspace{1cm} q(t)\equiv -\dfrac{1}{aH^2}\dfrac{d^2a}{dt^2}\ ,  \label{eq:H&q} \\
&j(t) \equiv \dfrac{1}{aH^3}\dfrac{d^3a}{dt^3} \ , \hspace{0.5cm}  s(t)\equiv\dfrac{1}{aH^4}\dfrac{d^4a}{dt^4}\ . \label{eq:j&s}
\end{align}
\end{subequations}
\begin{figure}[t]
\begin{center}
\includegraphics[width=0.8\textwidth]{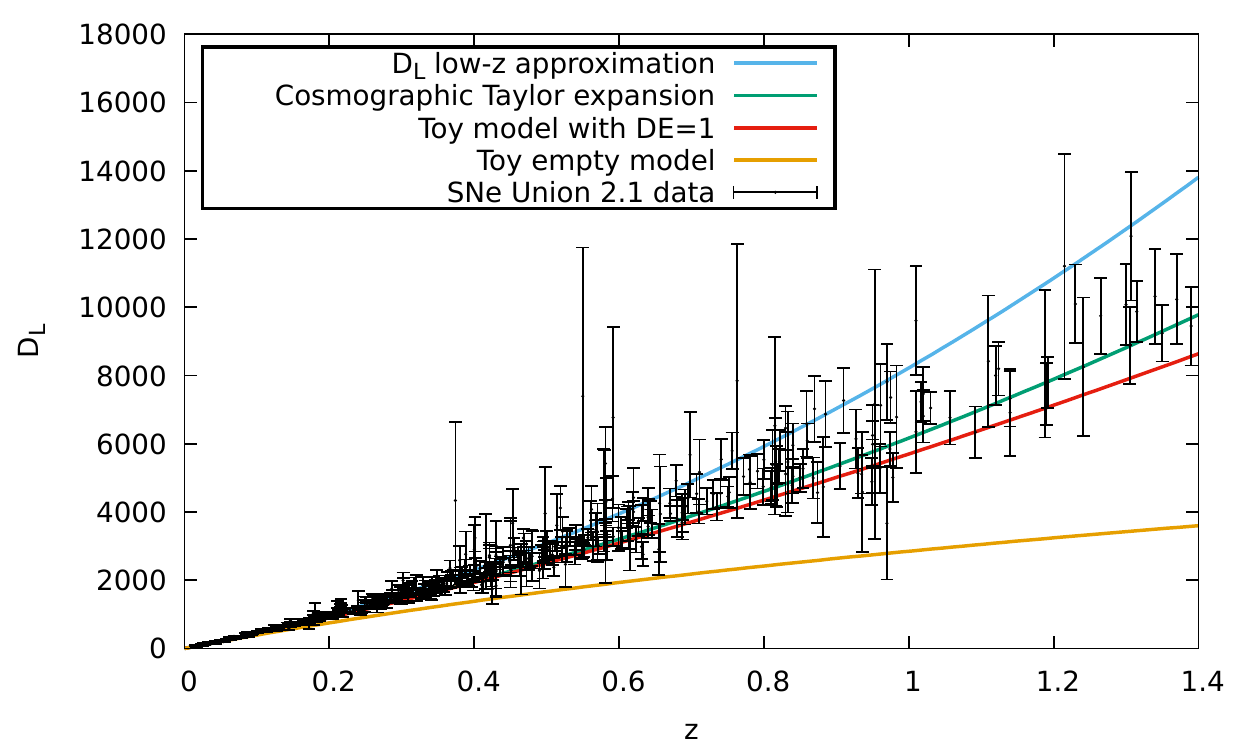}
\caption{Luminosity distance in terms of  redshift in the case of  low-redshift approximation of Eq.(\ref{eq:DL_lowz}) and the Taylor expansion of Eq.(\ref{eq:luminosity distance}) truncated at first order, with $q_0=-1/2$, in comparison with two toy models and data from Supernova Cosmology Project ``Union2.1" SNeIa compilation \cite{Amanullah:2010vv, Suzuki:2011hu}}
\label{fig:dL_1}
\end{center}
\end{figure}
These parameters give  information on the late dynamics of the universe. Indeed, the sign of the deceleration parameter indicates whether the universe is accelerating or decelerating; the sign of $j$ determines the change of the universe dynamics, and the value of $s$ is necessary to discriminate between  evolving dark energy  or  cosmological constant.

In other words,  we can rewrite the luminosity distance relation using the definition of Eq.(\ref{eq:z_a}) and Eq.(\ref{eq:scale factor}) and obtaining the Taylor expansion

\begin{align}
D_L(z)=&\ \dfrac{zc}{H_0}\bigg[1+\dfrac{z}{2}(1-q_0) -\dfrac{z^2}{6}\left(1-q_0-3q_0^2+j_0\right)+ \nonumber \\
&\hspace{0.8cm}+\dfrac{z^3}{24}\left(2-2q_0-15q_0^2-15q_0^3+5j_0+10q_0j_0+s_0\right)+\mathcal{O}(z^4)\bigg].
\label{eq:luminosity distance}
\end{align}
The truncation at low  orders of the above result is plotted with green solid line in Fig. \ref{fig:dL_1}, where it is compared with the low-$z$ luminosity distance approximation of Eq.(\ref{eq:DL_lowz}) (light blue curve) and also with two toy models: the first assumes that  the cosmological constant, $\Omega_{\Lambda}$, fully drives the universe evolution (i.e.  $\Omega_{\Lambda} =1$ and the matter density of  the universe is  $\Omega_m =0$), and a {\textit{empty Milne model}},  that assume  zero energy density and  negative spatial curvature.
We can see that the behaviors of the different approaches are all compatible up to $z \sim 0.2$.

It is worth noticing that the restricted convergence of the Taylor series, as well as the  arbitrary order of truncation of the series, make this method poorly predictive for cosmographic analysis for  $z > 1$.
Also, the degeneracy among the cosmographic coefficients of the Taylor series give rise to the  impossibility to measure them separately. In fact, it is  only the sum of them that leads to different results depending on the probability distribution associated with each coefficient. 
These problems can be partially alleviated by adopting the Pad\'e rational polynomials  \cite{Baker66} to construct a  cosmographic series based only on the assumption of cosmic homogeneity and isotropy (see  Ref.\cite{Capozziello:2019cav} for details). The Pad\'e polynomial approach to cosmography will be considered in the next section.

\section{The Pad\'e Polynomials Cosmography and $f(z)$CDM approach}
\label{Sec:Pade}
The Pad\'e Approximation for cosmographic analysis \cite{Gruber:2013wua, Wei:2013jya, Capozziello:2019cav, Rezaei:2017yyj, Capozziello:2018jya, Dutta:2018vmq,  basilakos, Aviles:2014rma, Capozziello:2017ddd} showed to have larger convergence radius than the Taylor series expansion and it is proving to be a reliable choice to extent the analysis to higher redshifts.
In this context, the approximate luminosity distance depends on the Pad\'e approximant $P_{(n,m)}$, of order $n/m$, that is the approximation of the ratio between two standard Taylor series of a generic function $f(z)=\sum_{i=0}^\infty c_iz^i$, 
\begin{equation}
P_{n,m}(z)=\dfrac{\displaystyle{\sum_{i=0}^{n}a_i z^i}}{1+\displaystyle{\sum_{j=1}^{m}b_j z^j}}\,.
\label{eq:Pnm_Pade}
\end{equation}
A detailed study of Pad\'e approximation in the cosmographic context is reported in Ref.\cite{Aviles:2014rma,Capozziello:2019cav}, where also the full expression of the coefficients of Eq.(\ref{eq:Pnm_Pade}) are provided.
The behaviour of the Hubble parameter evolution $H(z)$, using Pad\'e series has been tested at low redshift, finding that it is a reliable choice to extent the analysis up to $z \sim 6$ for a Pad\'e approximation of order (2,2) \cite{Capozziello:2018jya, Dutta:2018vmq}.
In this latter case, the background evolution reads like \cite{Capozziello:2018jya}

\begin{equation}
\label{eq:H_DE}
E(z) = \frac{H(z)}{H_{0}} = \frac{P_{0}+P_{1}z+P_{2}z^{2}}{1+Q_{1}z+Q_{2}z^{2}}.
\end{equation}
where

\begin{equation}\nonumber
P_0 =  1,
\end{equation}
\begin{equation}\nonumber
P_1 =  H_1 + Q_1,
\end{equation}
\begin{equation}\nonumber
P_2 = \frac{H_2}{2} + Q_1H_1 +Q_2,
\end{equation}
\begin{equation}\nonumber
Q_1 = \frac{-6 H_1 H_4 +12 H_2 H_3}{24 H_1 H_3 - 36 H_2^2},
\end{equation}
\begin{equation}
\label{eq:P_Q}
Q_2 = \frac{3 H_2 H_4 -4 H_3^2}{24 H_1 H_3 - 36 H_2^2},
\end{equation}
and  $H_{1}, H_{2}, H_{3}$ and $H_{4}$ are related to cosmographic parameters ($q_{0}, j_{0}, s_{0}, l_{0}$) as

\begin{equation}\nonumber
H_{1}=H_{10}/H_{0}= 1+q_{0}
\end{equation}
\begin{equation}\nonumber
H_{2}=H_{20}/H_{0}= -q_{0}^{2}+j_{0}\,,
\end{equation}
\begin{equation}\nonumber
H_{3}=H_{30}/H_{0}= 3q_{0}^{2}(1+q_{0})-j_{0}(3+4q_{0})-s_{0}\,,
\end{equation}
\begin{equation}
\label{eq:Hi}
H_{4}=H_{40}/H_{0}= -3q_{0}^{2}(4+8q_{0}+5q_{0}^{2}) +j_{0}(12+32q_{0}+25q_{0}^{2}-4j_{0})+ s_{0}(8+7q_{0})+l_{0}\,.
\end{equation}
\begin{figure}
\begin{center}
\includegraphics[width=0.8\textwidth]{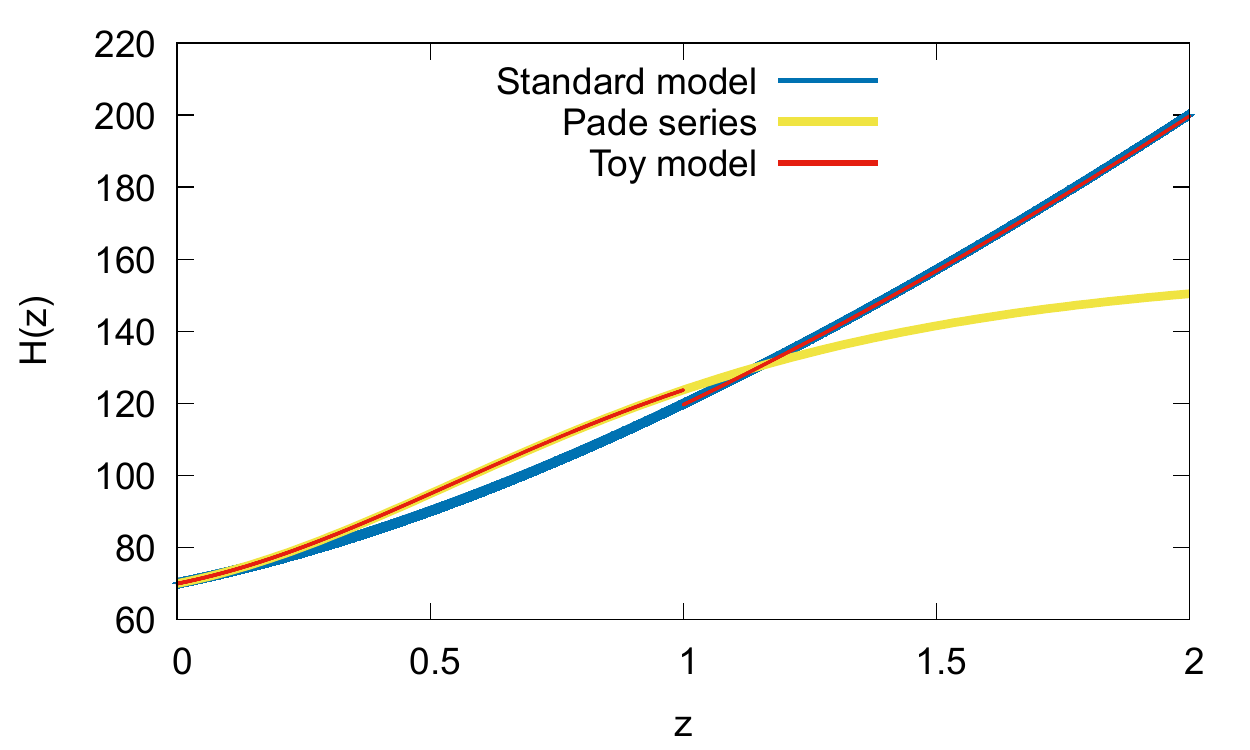}
\caption{$H(z)$ evolution expansion of a toy model using Eq.(\ref{eq:H_DE}) up to $z=1$ and the minimal $\Lambda$CDM model afterwards. For this model (red line) we assume  $q_0=-0.6$, $j_0=1.97$ and $s_0=l_0=0$. 
For comparison, we also show the minimal $\Lambda$CDM model (blue curve) and cosmographic Pad\'e $H(z)$ series (yellow curve), ever using the same choice of cosmographic parameters values.}
\label{fig:H_z_splitted}
\end{center}
\end{figure}
Although the above approach, and the cosmography in general, is  interesting for studying the evolution of the universe at large scales and late times, its independence of the model  falls when it is extended to higher redshifts. Despite the fact that series convergence  can be improved with reliable choices, getting closer to redshifts where the cosmological constant is no longer dominant,  we must necessarily assume a cosmological model in order to take into account the evolution of cosmic fluids. In particular, it is needed the value of  current energy density parameter associated to the non-relativistic matter, $\Omega_m$, which is well constrained using the CMBR data.
If, on one hand,  $\Omega_m$ value can be assumed ``a priori", using constraints of the standard cosmological model, on the other hand, the generality of  cosmographic results are biased because conclusions can be drawn only considering specific values of cosmological parameters. In other words, there is a circularity problem because the \textit{cosmographic 
model-independent approach} becomes \textit{model dependent}. 

\begin{figure}
\begin{center}
\includegraphics[width=0.7\textwidth]{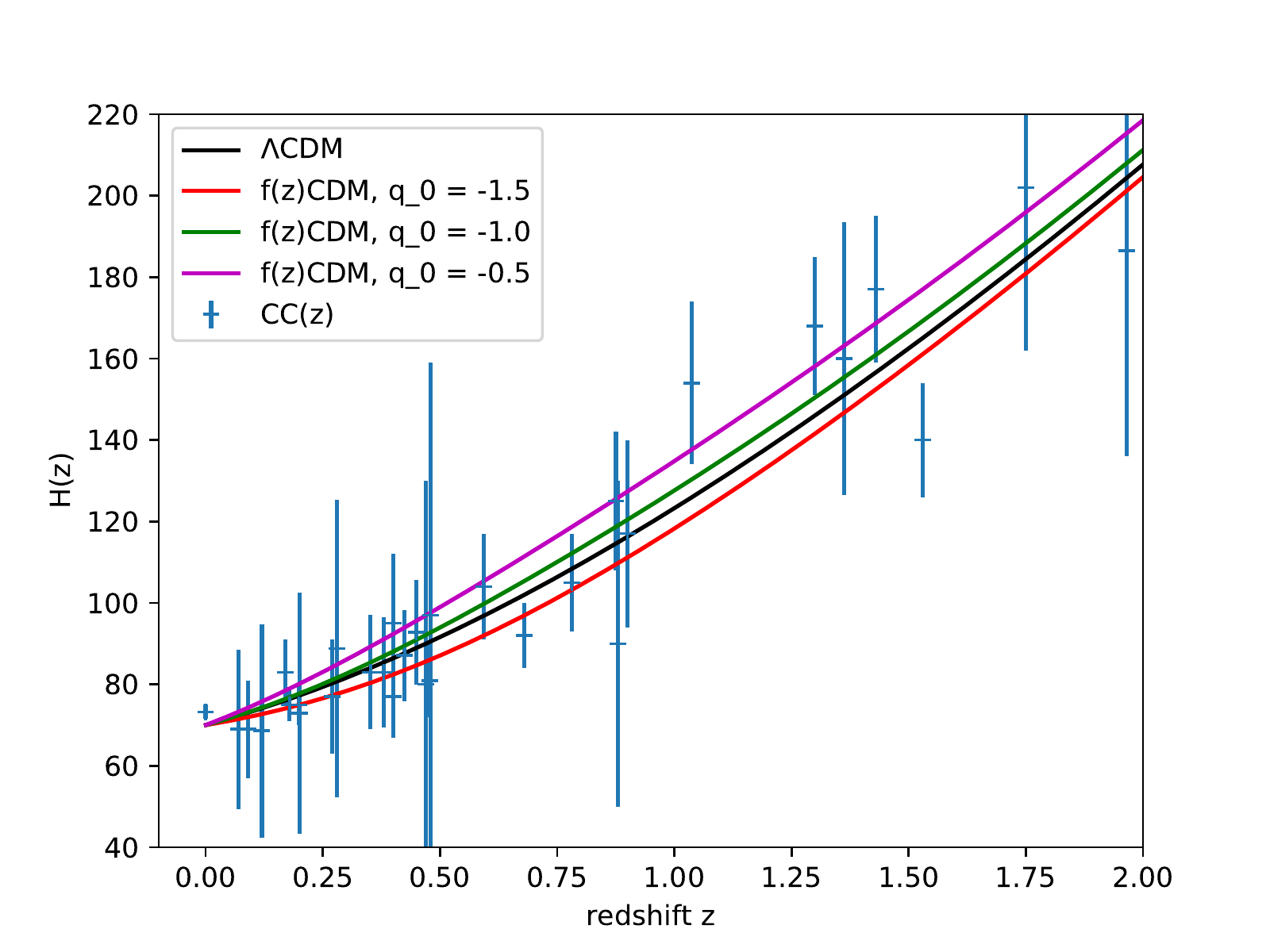}
\caption{$H(z)$ evolution expansion of $f(z)$CDM model, where $j_0$ is fixed to $1.97$ and $s_0=l_0=0$, while different values of deceleration  parameter are assumed, as  $q_0=-0.5$ (magenta curve), $q_0=-1$ (green curve), $q_0=-1.5$ (red curve). For comparison, the minimal $\Lambda$CDM model is drawn with a black solid line and the Cosmic Clock data of Tab.\ref{tab:Hz} are reported. In this plot,  we assume $\Omega_m = 0.3$, $\Omega_{\Lambda} = 1-\Omega_m$ and $H_0 = 70$ Km/s/Mpc.}
\label{fig:H_z_fzCDM}
\end{center}
\end{figure}
A reasonable way  to link the cosmographic parameterization with cosmological models, without assuming  ``a priori" values, can be to build up a cosmological model considering the standard $\Lambda$CDM evolution up to an appropriate redshift, i.e. the scale of equivalence matter-$\Lambda$, and then a cosmographic  approximation (for example based on Pad\'e polynomials as Eq.(\ref{eq:H_DE})), to describe the universe kinematic at lower redshift. In Fig. \ref{fig:H_z_splitted}, a comparison of this toy model (red line) for an arbitrary choice of redshift of transition (i.e. $z=1$) and cosmographic parameters values is shown
\footnote{Here we assume the best fit parameters values in Ref. \cite{Capozziello:2018jya}.}
 for the minimal $\Lambda$CDM evolution and cosmographic Pad\'e $H(z)$ series. This toy model curve overlaps the Pad\'e one for $z<1$ and the $\Lambda$CDM curve for $z>1$, presenting a step in the redshift where the two model are linked at  $z=1$. 

To prevent such a discontinuity, and also to avoid choosing the redshift value where the modeling step takes place, we can replace the $\Lambda$CDM model with a $f(z)$CDM model, where $\Lambda$ is substituted by a Pad\'e cosmographic series,  i.e. considering the $H(z)$ evolution as

\begin{equation}
\label{eq:H_full}
E(z)^2 = \left(\frac{H(z)}{H_{0}}\right)^2 = \Omega_k (1+z)^2 + \Omega_m (1+z)^3 + \Omega_r (1+z)^4 + \Omega_f f(z)\,,
\end{equation}
with, in the specific case, $f(z)$ is given by 
\begin{equation}
\label{eq:fz}
f(z) = \frac{P_{0}+P_{1}z+P_{2}z^{2}}{1+Q_{1}z+Q_{2}z^{2}} \,.
\end{equation}
The entries above are the spatial curvature density $\Omega_k$, the matter density (baryonic and cold dark matter (CDM) component of standard cosmology) $\Omega_m$, and the radiation $\Omega_r$. 
Hereafter, we will choose for the sake of simplicity $\Omega_k=0$, so that $\Omega_m + \Omega_r + \Omega_f = 1$.
The term $\Omega_{f}$ is introduced to describe the current universe evolution after a given polynomial cosmography evolving with redshift $z$ is adopted. 
In Fig. \ref{fig:H_z_fzCDM} we show the $f(z)$CDM behavior for different deceleration parameter values, comparing with the minimal $\Lambda$CDM model (black line), i.e. the case in which $f(z) \rightarrow \Lambda$. 
It is worth noticing that also the  cosmological equation of state can be reconstructed according to a similar  method \cite{Escamilla1} and that a deep machine learning approach to cosmological models can be constructed according to iterative procedures tracking the evolution of dark energy models \cite{Escamilla2}. Also in that case, a $f(z)$CDM procedure can be implemented.
\begin{figure}
\begin{center}
\centering
\includegraphics[width=0.65\textwidth]{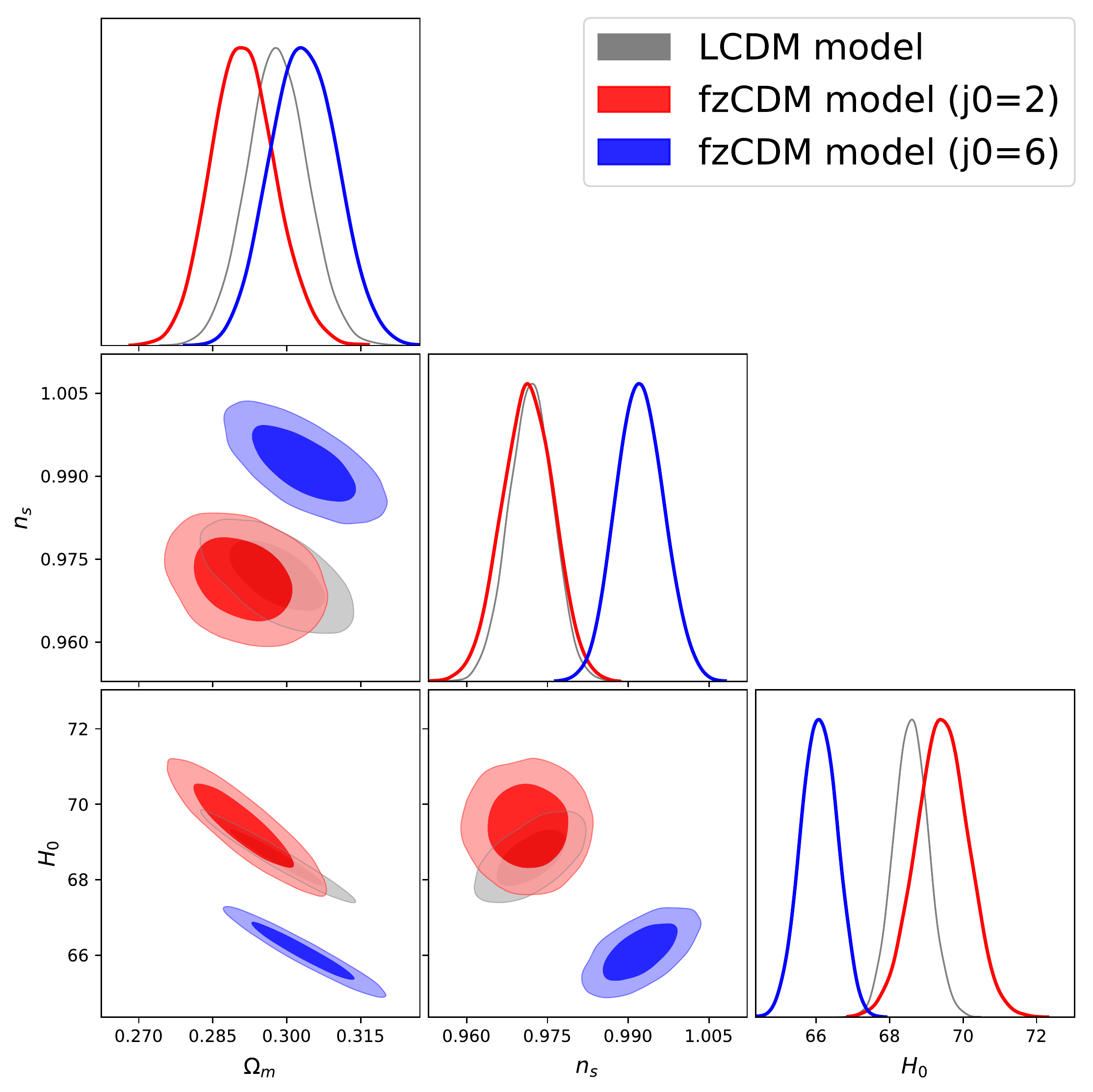}
\includegraphics[width=0.25\textwidth]{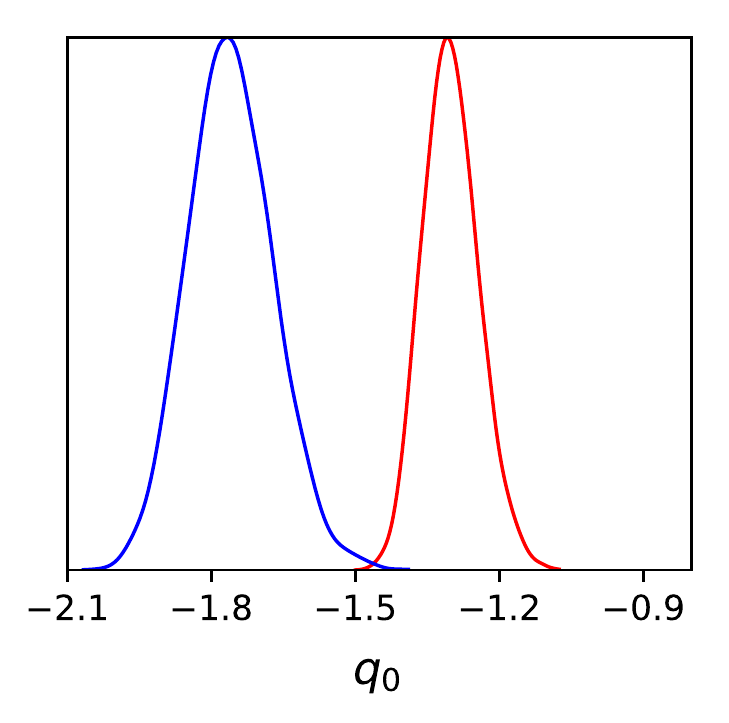}
\caption{C.L of f(z)CDM model with one parameter extension using base-2015 dataset. The deceleration  parameter is free to vary, while $j_0$ is fixed in 2 (red line) and 6 (blue line), and $s_0=l_0=0$.}
\label{fig:test_1par}
\end{center}
\end{figure}
\section{The Method}
\label{Sec:Analysis}

Let us introduce now  the codes used for our analysis, and also the main observational data set we considered to perform the parameters estimation.

In order to compute the theoretical predictions of the $f(z)$CDM model, among which the CMBR temperature power spectrum, we modified the current version of the Code for Anisotropies in the Microwave Background ({\sc CAMB})~\cite{camb} where the background evolution is now given by Eqs.~(\ref{eq:H_full})-(\ref{eq:fz}), also using Eqs.(\ref{eq:P_Q})-(\ref{eq:Hi}). 
We performed the parameter constraints using a Monte Carlo Markov chain statistical analysis, modifying the available parameter estimation packages {\sc CosmoMC}~\cite{cosmomc} to our purpose and analyzing both the $f(z)$CDM and the $\Lambda$CDM model for a given range of values of the cosmological and cosmographic parameters. 

We choose to build three datasets for our analysis. The first, called ``base-2015" in the next sections, combining

\begin{itemize}

\item Cosmic Microwave Background Radiation (CMBR) measurements, through the Planck (2015) data \cite{Adam:2015rua, Aghanim:2015xee} that include both temperature power spectra (TT) over the range $\ell \in [2 - 2508]$ and low-$\ell$ (2 - 29) temperature-polarization cross-correlation likelihood;

\item Baryon Acoustic Oscillation (BAO) distance measurements, using 6dFGS ~\cite{Beutler:2011hx}, SDSS-MGS~\cite{Ross:2014qpa}, and 
BOSS DR12~\cite{Alam:2016hwk} surveys, as considered by the Planck collaboration;

\item Supernovae Type Ia (Pantheon sample), that is the latest compilation of 1048 data points, covering the redshift range $[0.01 : 2.3]$~\cite{Scolnic:2017caz};

\item Hubble constant of latest Riess (2019) work (R19), $H_0$ = $74.03 \pm 1.42$ km/s/Mpc ~\cite{Riess:2019cxk}, that is in tension at 4.4$\sigma$ with CMBR estimation within the minimal cosmological model. This measurement is implemented by default in the package {\sc CosmoMC} by imposing a Gaussian prior for the Hubble parameter constraint.

\item Cosmic Clock (CC) data, as measurements got from differential
age treatment as showed in Tab.\ref{tab:Hz}, in which H(z) is given by km/s/Mpc.

\end{itemize}
Furthermore,  in the second  data set, we replace the Planck 2015 data with Planck 2018 \cite{Aghanim:2019ame}, using TT CMBR power spectra, and HFI polarization EE likelihood  at $\ell \leq 29$. 
We refer to the new base dataset as Planck 2018 + BAO + Pantheon + R19 + CC as ``base-2018", to recall that here we use the Planck release 2018.

Finally, we  include also the Dark Energy Survey (DES) data, considering both galaxy clustering and cosmic shear measurements from the combined probe Year 1 results \cite{Troxel:2017xyo, Abbott:2017wau,Krause:2017ekm} with the ``base 2018". We refer to this extended dataset as ``base-2018 + DES".
\begin{figure}
\begin{center}
\includegraphics[width=0.7\textwidth]{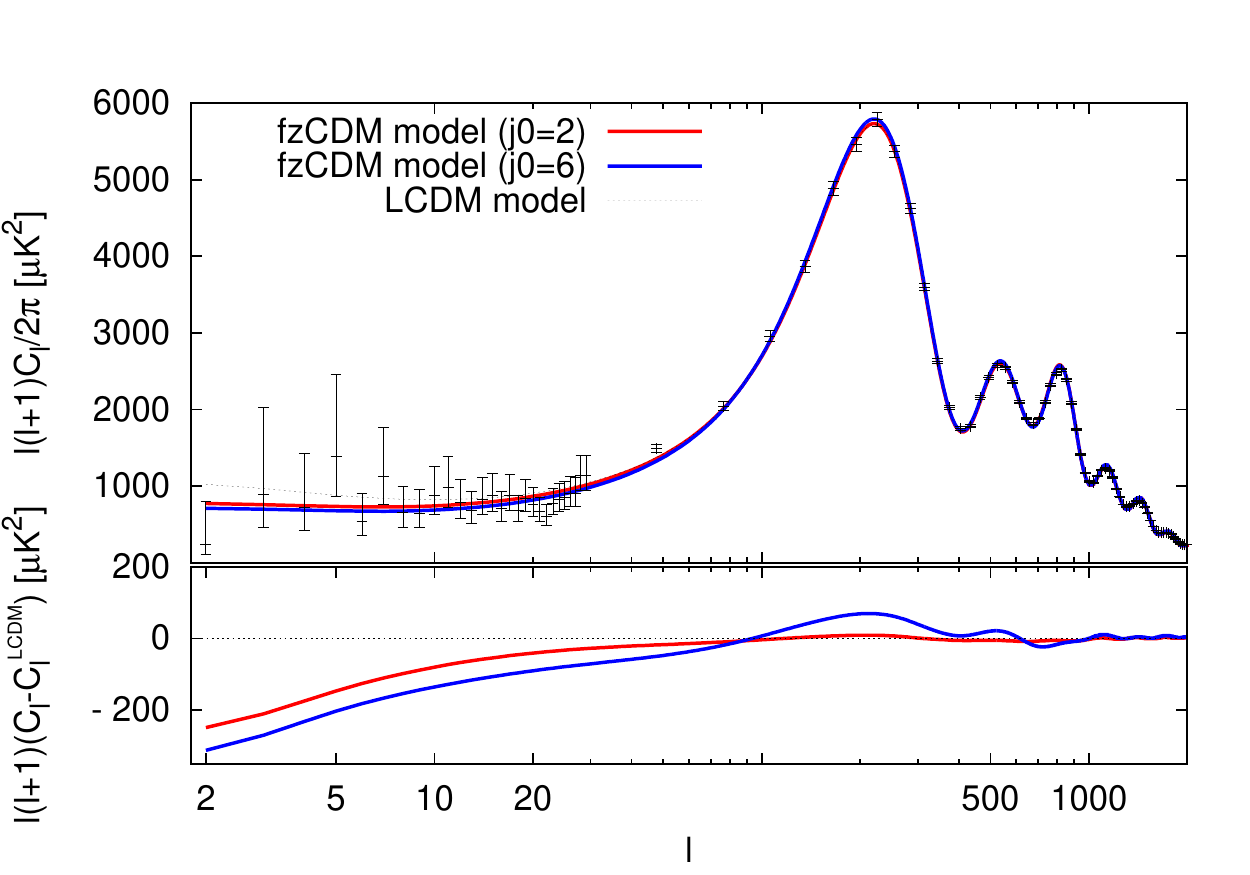}
\caption{Temperature Anisotropy power spectrum for the best fit f(z)CDM model, with one parameter extension, constrained with base-2015 joint data. Superimpose data are from Planck 2015 release. }
\label{fig:TT}
\end{center}
\end{figure}
\section{Results}
\label{Sec:Results}
In this section, we present the results,  showing how our analysis  increases the  truncation orders of cosmographic series. The aim is to test the $f(z)$CDM model taking care to understand the impact of different truncation orders on the parameter constraints.

We firstly analyze the case of $f(z)$CDM model with {\it{one parameter  extension}}, i.e. with a free deceleration   parameter and fixed $j_0$, $s_0$, $l_0$, using the base-2015 data. In Fig.\ref{fig:test_1par} we show our  results for two different (and arbitrary) $j_0$ parameter choices, i.e. assuming $j_0=2$ (red curve) and considering $j_0=6$ (blue curve). For comparison, we also show the C.L. curves of the minimal $\Lambda$CDM model using the same dataset (left panel). We can see that the $f(z)$CDM model with $j_0=2$ is able to constrain cosmological parameters values compatible with the standard model ones, even allowing for slighty higher $H_0$ at the cost of lower values of matter density $\Omega_{m}$. Let we stress that $\Omega_{m}$ is degenerate with the cosmographyc parameters, see also Tab.\ref{tab:Pade}, inferring the weight of the approximation made when it is fixed to an arbitrary value when analyzing the cosmographic models.
Noteworthy, the positive correlation between the $H_{0}$ parameter with the spectral index noticed in the standard model (and widely explored in the literature, see Refs. \cite{Gerbino:2016sgw, Benetti:2017gvm, Benetti:2017juy} and references within) vanishes for the  $f(z)$CDM model with $j_0 = 2$. 
The $q_0$ parameter is well constrained by the used dataset, with probability posterior distribution indicated in the right panel of Fig.\ref{fig:test_1par}. 
The temperature anisotropy power spectra, using best fit values of our analysis, are plotted in Fig. \ref{fig:TT}, where we can see the lack of power at low multipoles for these models, if compared to  $\Lambda$CDM, due to the late-time integrated Sachs-Wolfe effect \cite{Ade:2015rim}. In other words, the CMBR is sensitive to the cosmographyc parameters.

Then, we can analyze the $f(z)$CDM model with \textit{two parameters  extension}, i.e. with a free deceleration and jerk parameters and fixed $s_0=l_0=0$, using the base-2015 data. 
Looking at Fig. \ref{fig:test_2par_3par} (green curve), the probability posterior distribution of $j_0$ (right panel) show that it prefers values lower than those assumed in the previous analysis, i.e. $j_0=2$, thus implying a constraint of higher $q_0$ values. This is in agreement with previous results \cite{Capozziello:2019cav,Capozziello:2017nbu,Capozziello:2018jya}, also allowing for higher $H_0$ mean value with respect to the standard model, compatible in $1\sigma$ with this latter. Also, we show, in Fig. \ref{fig:test_2par}, the C.L. of cosmographic and matter density parameters. Besides the degeneration between cosmographic parameters, we note the correlation in $\Omega_m - q_0$ plane and the anti-correlation between the cold dark matter density and the jerk parameter. 

Finally, using the base-2015 data we also leave the \textit{$s_0$ parameter as a new free parameter} of the model (see Fig. \ref{fig:test_2par_3par}, yellow lines). We can see that the snap posterior values is fully compatible with zero, i.e. $s_0 = -0.1 \pm 0.6$ in $1\sigma$, and also the constraints on the cosmological and cosmographic parameters do not change with respect to the case where $s_0$ was fixed to zero (green line). Noteworthy, the $s_0$ parameter constrained is in fully agreement with the previous results \cite{Capozziello:2019cav}, showing a worse constraint than the other cosmographic parameters.  
\begin{figure}
\begin{center}
\includegraphics[width=0.65\textwidth]{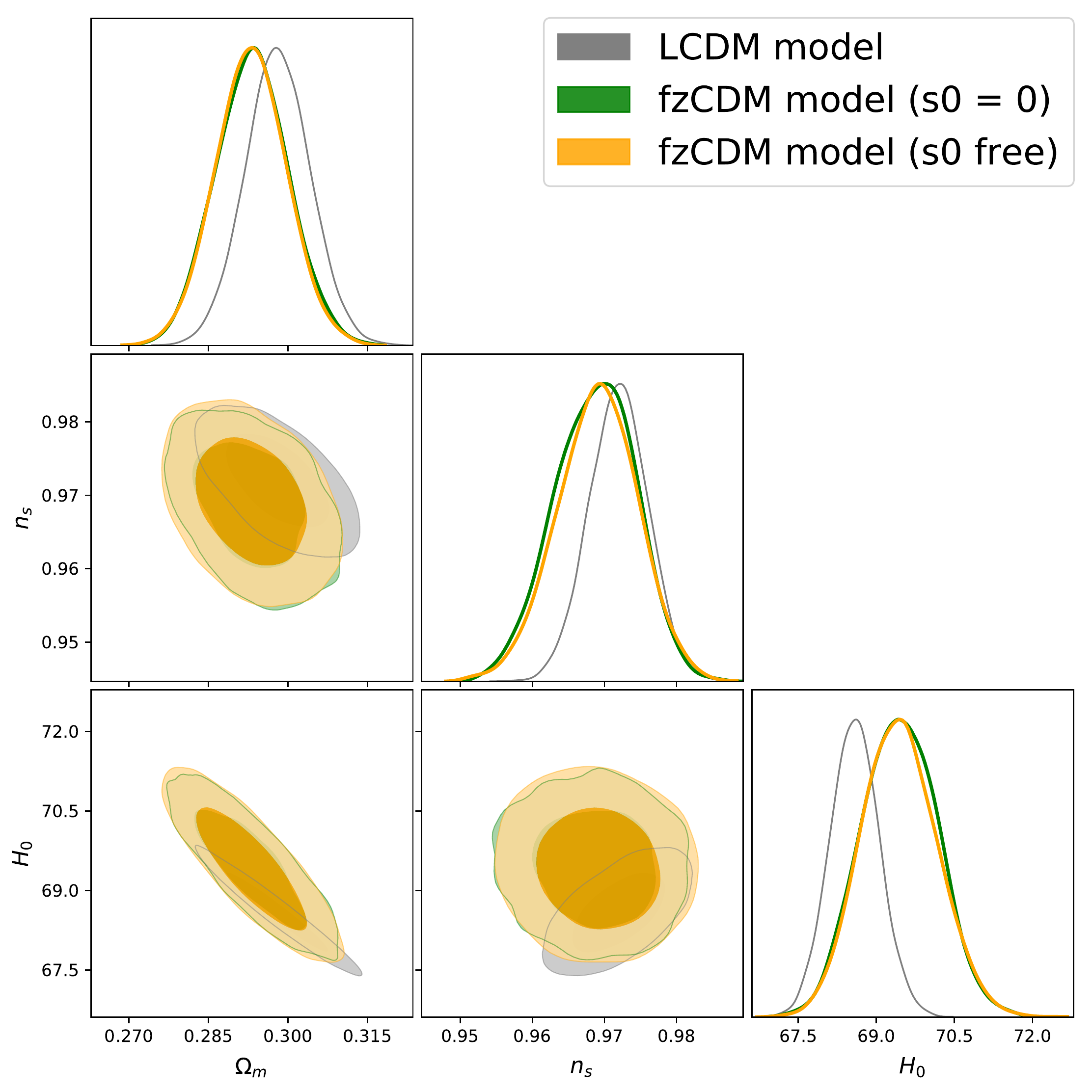}
\includegraphics[width=0.25\textwidth]{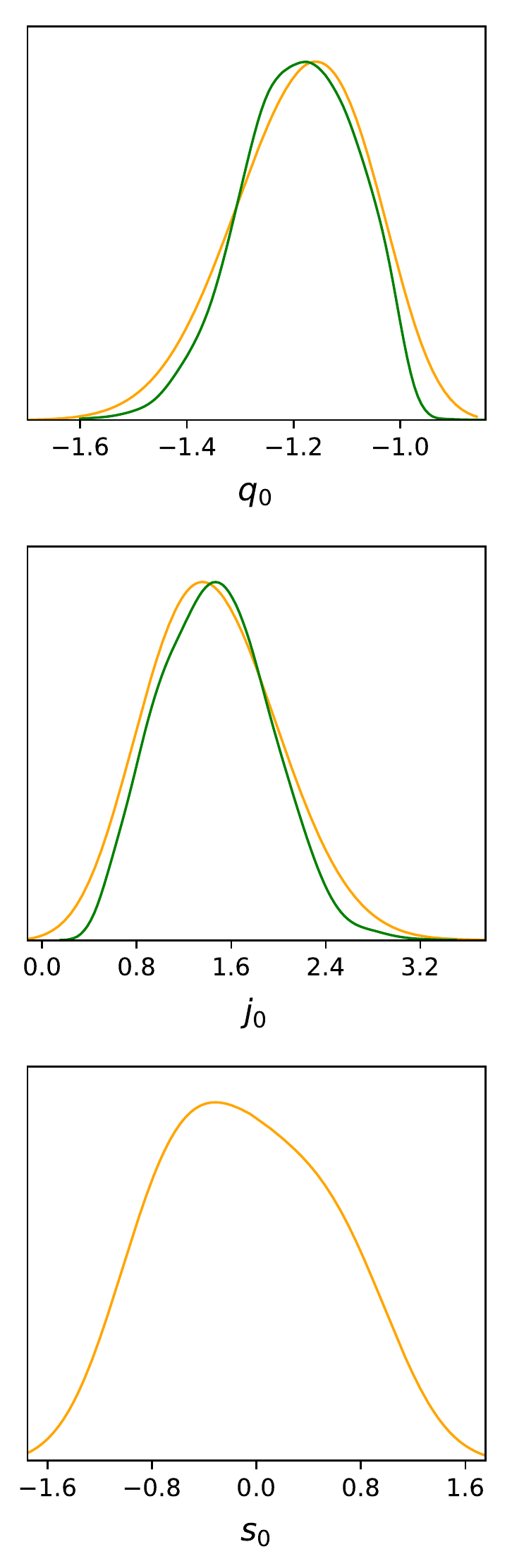}
\caption{C.L of $f(z)$CDM model with two parameters extension (green line) and three parameters extension (yellow line), using the base-2015 data. The deceleration and jerk parameters are free to vary, while $s_0$ is fixed in the case of the green curves and free for the yellow curves. In any case, $l_0=0$.}
\label{fig:test_2par_3par}
\end{center}
\end{figure}
\begin{figure}
\begin{center}
\includegraphics[width=0.80\textwidth]{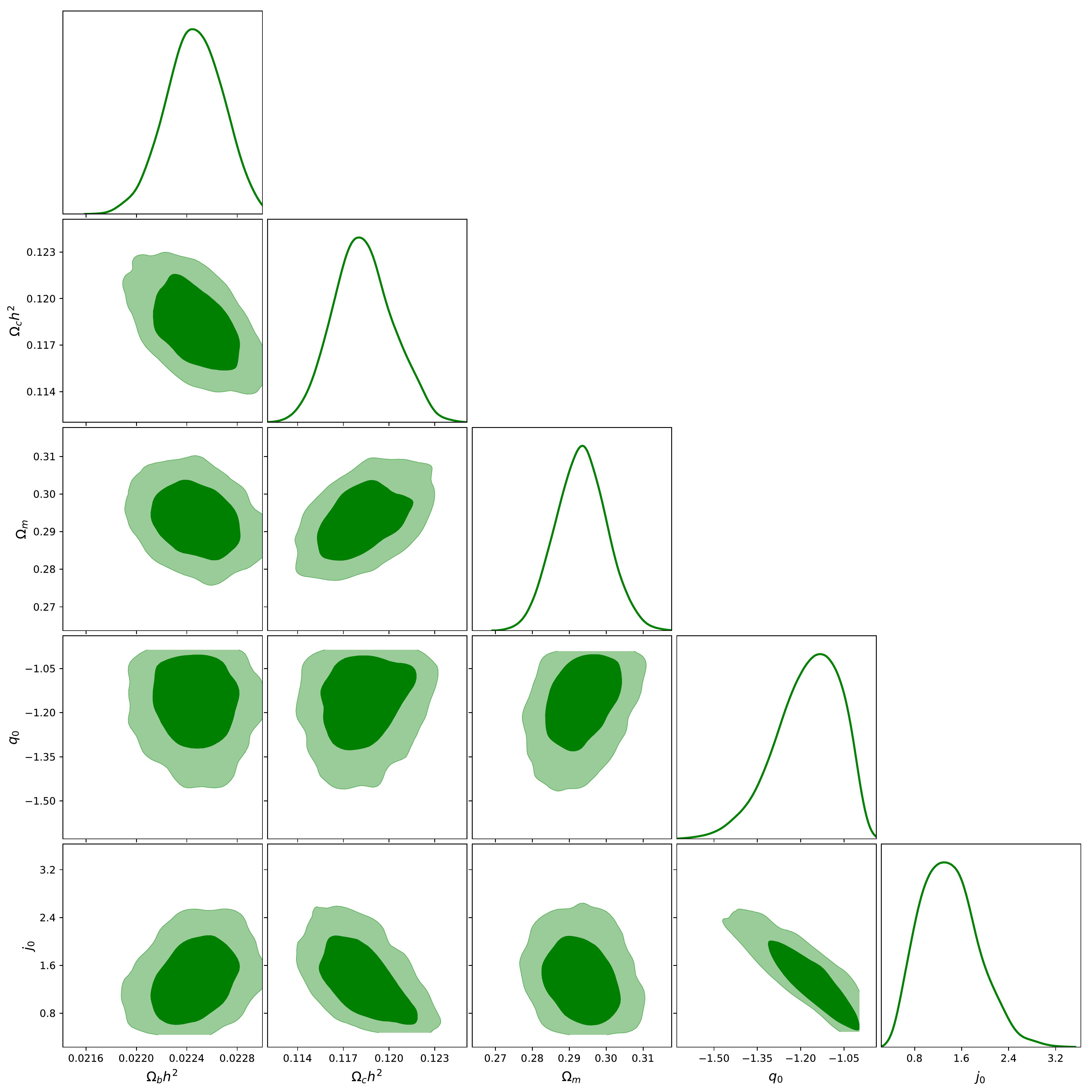}
\caption{C.L of $f(z)$CDM model with two parameters extension, using the base-2015 data. The deceleration and jerk parameters are free to vary, while $s_0=l_0=0$.}
\label{fig:test_2par}
\end{center}
\end{figure}
\begin{table}[]
\centering
\caption{{
$68\%$ confidence limits for the model parameters constrained using base-2015 dataset .
The $\Delta \chi^2_{best} = \Delta \chi^2_{\Lambda CDM} - \Delta \chi^2_{model}$ refers to the best fit of the model (negative value means a better $\chi^2$ of the reference model, $\Lambda$CDM .}
\label{tab:Pade}}
%\scalebox{1.3}{
\begin{tabular}{|c|c|c|c|c|}
\hline
{Parameter}&
{\textbf{$\Lambda$CDM }}&
{\textbf{$f(z)$CDM ($j_0= 2$)}}&
{\textbf{$f(z)$CDM ($j_0 = 6$)}}&
{\textbf{$f(z)$CDM ($j_0$ free)}}
\\
\hline
$100\,\Omega_b h^2$ 	
& $2.245 \pm 0.019$ % 
& $2.252 \pm 0.021$ % 
& $2.311 \pm 0.020$ % 
& $2.246 \pm 0.022$ % 
\\
$\Omega_{c} h^2$	
& $0.1171 \pm 0.0011$ % 
& $0.1172 \pm 0.0015$ % 
& $0.1087 \pm 0.0012$ % 
& $0.1182 \pm 0.0019$ 
\\
$\Omega_{m}$	
& $0.2981 \pm 0.0061$ % 
& $0.2913 \pm 0.0064$ % 
& $0.3035 \pm 0.0066$ % 
& $0.2929 \pm 0.0067$ 
\\
$H_0$
& $ 68.58 \pm 0.47 $ 
& $ 69.42 \pm 0.71 $ 
& $ 66.08 \pm 0.48 $ 
& $ 69.46 \pm 0.71 $ %26
\\
$q_0$
& - 
& $ -1.30 \pm 0.06$  
& $ -1.75 \pm 0.09$  
& $ -1.19 \pm 0.10$ %32
\\
$j_0$
& - 
& 2
& 4  
& $ 1.5 \pm 0.5$ %32
\\
\hline
\hline
$\Delta \chi^2_{\rm best}$         
& $ - $ % 
& $5,4$ % 
& $ -51,6$ %
& $6,6$ % 
\\
\hline
\end{tabular}%}
\end{table} 

Considering now the dataset base-2018 and base-2018+DES introduced in  previous section, we focus on the $f(z)$CDM model with three parameter extension. 
Replacing the new Planck release (2018) with the previous one (2015), we obtain the magenta curves shown in Fig. \ref{fig:test_3par_2018}, while the yellow ones, referring to the base-2015 analysis,  is  already presented above. With cyan curve, we also plot the base-2018 + DES data. We note that the new Planck release prefers slightly lower values of $q_0$ parameter with respect the 2015 TT likelihood, that implies an higher value of the jerk parameter. 
The addition of DES data constrains lower value of cold dark matter density, $\Omega_c$, and prefers slightly negative values of the snap parameter. The cosmographic parameter $1\sigma$ mean values are presented in Tab. \ref{tab:3_parameter}.
 \begin{figure}
\begin{center}
\includegraphics[width=0.75\textwidth]{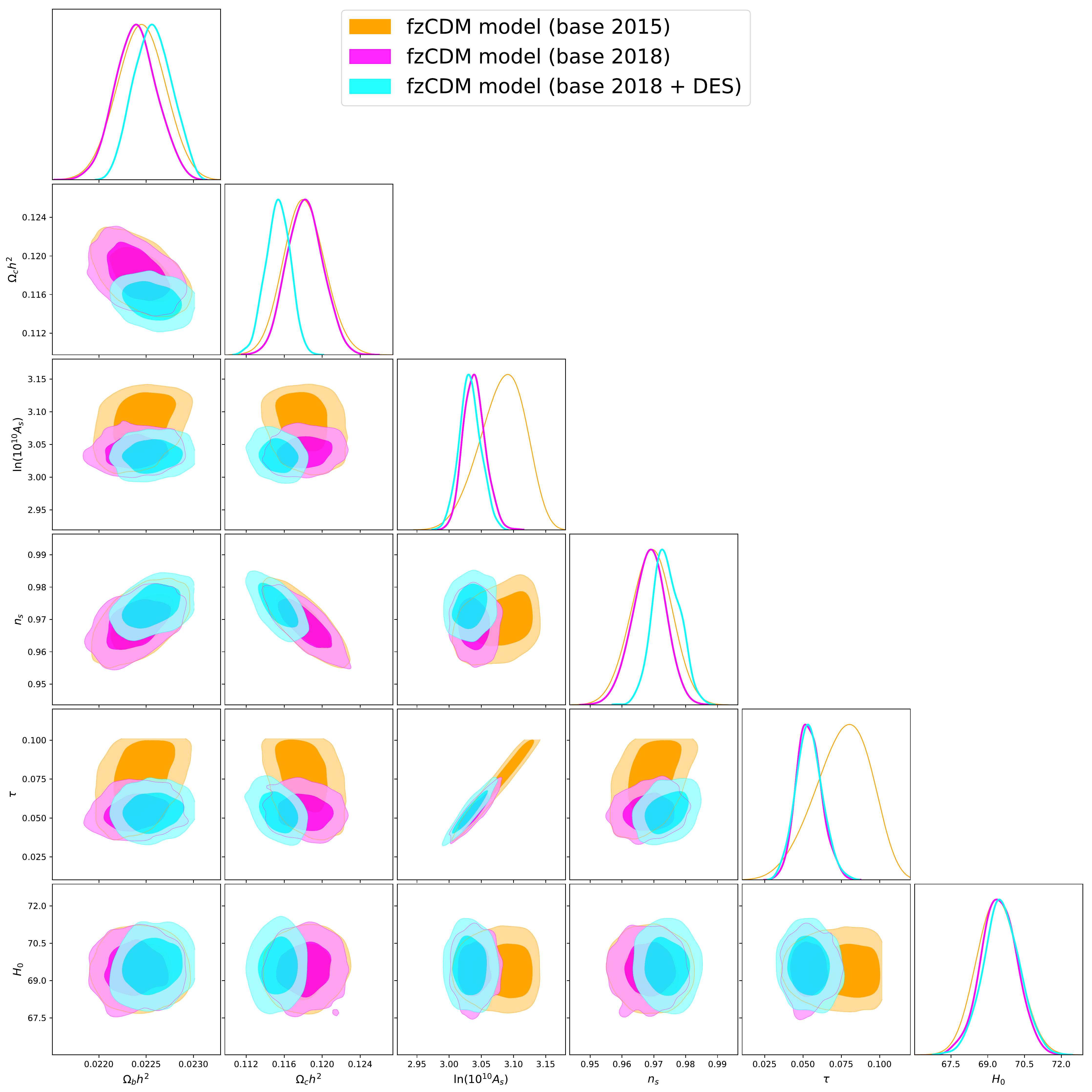}
\includegraphics[width=0.2\textwidth]{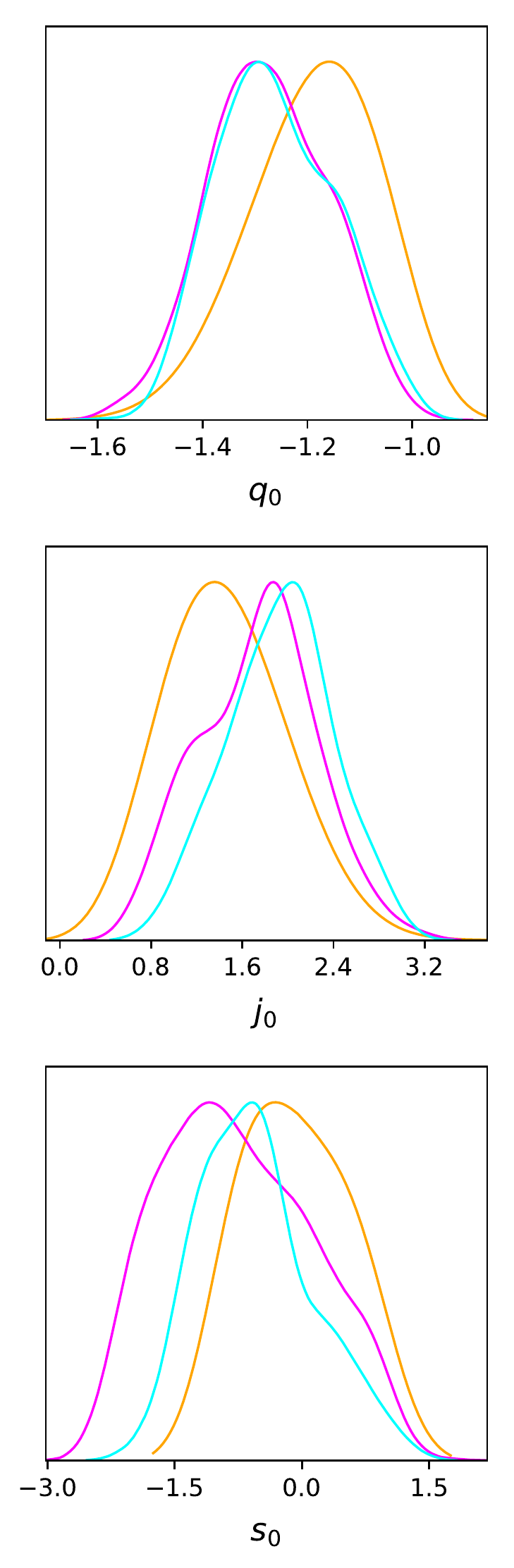}
\caption{C.L of $f(z)$CDM model with three parameters extension using base-2015 data (orange line), base-2018 data (magenta line) and base-2018 + DES data (cyan line).}
\label{fig:test_3par_2018}
\end{center}
\end{figure}        
\begin{table}[!]
\centering
\caption{{$68\%$ confidence limits for the cosmographic model parameters of the $f(z)$CDM model with three parameters extension constrained using base-2015, base-2018 and base-2018 + DES dataset.
}
\label{tab:3_parameter}}
%\scalebox{1.4}{
\begin{tabular}{|c|c|c|c|}
\hline
{Parameter}&
{base-2015}&
{base-2018}&
{base-2018 + DES}
\\
\hline
$q_0$
& $ -1.18 \pm 0.11$ 
& $ -1.27 \pm 0.11$ 
& $ -1.26 \pm 0.11$
\\
$j_0$
& $1.5 \pm 0.5$
& $1.7 \pm 0.5$ 
& $1.9 \pm 0.5$ 
\\
$s_0$
& $-0.1 \pm 0.6$
& $-1.1 \pm 0.9$ 
& $-0.6 \pm 0.7$ 
\\
\hline
\end{tabular}%}
\end{table}        
 
\section{Discussion and Perspectives}
\label{Sec:Conclusions}

In this paper we proposed a generalization of $\Lambda$CDM, the parametric  $f(z)$CDM model, as a new strategy to describe the current universe evolution  assuming  a model-independent cosmographic parameterization.  The  approach is aimed to describe  observations that are showing tensions with respect to $\Lambda$CDM.   Specifically, the procedure consists in  constructing  a function of the redshift $f(z)$, based on specific polynomials, capable of describing the evolution of the standard cosmological model at a given redshift  \textit{without} imposing any dark energy  model. In principle, the whole information should come from the cosmographic parameters, entries of the luminosity distance and  derived from combined data sets of observational data.   

Specifically, we explored  the  $f(z)$CDM model choosing a Pad\'e series $P(2,2)$, that shows a good convergence up to high redshift. 
We considered several series orders introducing, step by step, higher cosmographic parameters. 
Our analysis highlight the crucial role played  by the series convergence problem  in allowing the self-consistency of  cosmological model. In fact, if the series does not guarantee convergence at high redshifts,  undesirable contributions come out and any cosmological model results biased. 
The Pad\'e series $P(2,2)$ proved to be unstable for high values of cosmographic parameters, although widely reliable in the selected priors used in this work. It is therefore necessary to explore new possibilities, such as Pad\'e polinomials with different orders, which may allows for a more fast convergence of the series \cite{Li:2019qic}.
Alternatively, one can evaluate the possibility to   use other   redshift parameterizations  \cite{Capozziello:2017ddd,Aviles:2014rma,Aviles:2012ay} or to construct the function $f(z)$ with other rational polynomials like the Chebyshev ones  \cite{Capozziello:2019cav,Capozziello:2017nbu}.

According to the considered particular model, the analysis confirms the anti-correlation between the parameters $q_0$ and $j_0$, and between the deceleration parameter and $\Omega_m$.
Noteworthy, the TT spectrum is sensible to the values of  cosmographic parameters, therefore the CMBR has proved to be an important dataset to be used in cosmography.
Also, $H_0$ prefers higher values than the standard case, although always in $1\sigma$ from the values of the vanilla $\Lambda$CDM.
Finally, considering DES galaxy clustering data, we found a slightly negative $s_0$ parameter constraint. Let we stress that the sign of the snap parameter could give some indication on the nature of dark energy, i.e. it can distinguish between evolving dark energy or cosmological constant. In particular, the negative value constrained by the base-2018 dataset, $s_0 = -1.1 \pm 0.9$, excludes the zero value at $1\sigma$, indicating a preference for models with evolving dark energy. However, this preference falls when the joined dataset with DES is considered.

We note that the results presented in this work can be the starting point for a completely new interpretation of the current cosmological tensions. Indeed, while in literature it has been exhaustively demonstrated that the theory of General Relativity well describes the whole evolution of the universe until recent times, tensions emerges at low-redshifts. 
At the same time, cosmography, using a third-order truncation (i.e. using two cosmographic parameters like the deceleration and the jerk), seems to better describe the current observations, showing a better $\chi^2$ than about 7 points with respect to the minimal cosmological model. 
Furthermore, galaxy clustering and cosmic shear measurements seem to have some sensitivity with higher-order parameters, such as $s_0$, where the CMBR data seem to be non-sensitive.
These facts  could point out that  the emerged tensions could  not indicate any  new physics, nor  they can depend on systematic data errors as widely proposed in  literature, but rather show only the need to include higher corrections in the model without assuming $\Lambda$CDM at background level. 
In other words, $\Lambda$CDM could be nothing else but a coarse grained model that needs to be improved with higher order corrections in view to achieve a self-consistent interpretation of phenomenology. Clearly  such corrections have to be physically interpreted according to the data that should be reliable both at early  and late epochs. 
A possibility, widely studied in the literature, is to provide  higher order corrections related to the geometry of the universe (coming as  remnants of quantum fluctuations in early universe  \cite{Star1,Barr1,Birr1}). 
In this context, several Extended Theories of Gravity\footnote{The General Relativity is a second-order theory, and higher-order gravitational corrections of the Hilbert-Einstein Lagrangian density can be achieved in the so-called Extended Theories of Gravity \cite{Capozziello:2011et}. Such  higher-order corrections can be also described by a perfect fluid in the field equations \cite{Capozziello:2019qlt} } have  been  tested, addressing phenomenology ranging from the primordial   inflationary  expansion to the current accelerated behavior \cite{Capozziello:2011et, Nojiri:2017ncd, Capozziello:2019cav}. 
Among these, it is particularly promising the Starobinsky model \cite{Starobinsky:1980te,Starobinsky:2007hu} that, according to the recent CMBR data, is one of the best candidate to fit inflationary behavior with the  possibility to address also the recent dark energy behavior  \cite{Benetti:2019smr, Motohashi:2010tb}. In this perspective, $H_0$ tension and other shortcomings of present observational cosmology could be solved, in principle,  considering geometric corrections originated from primordial fluctuations and propagating up to infrared scales.

Present  results show, as a perspective, that  both  higher performing cosmographic series and  more precise data at low redshift should be considered. 
If currently high-redshift spectroscopic surveys allow for accurate data, covering up to $z \sim 2$, 
in the near future experiments like DESI \cite{Levi:2019ggs} will cover  very large redshift range using Emission Line Galaxies (ELGs), Luminous Red Galaxies (LRGs) up to quasars, while others, as Large Synoptic Survey Telescope (LSST) \cite{Jha:2019rog}, are proposed to improve such a redshift range with million of spectroscopic targets, i.e. Lyman-Break galaxies (LBGs) and Lyman-$\alpha$ emitters (LAEs), allowing even better measurements. At the same time, MegaMapper \cite{Schlegel:2019eqc} will provide a high-redshift spectroscopic survey from galaxy redshifts at $2<z<5$,  allowing for a better constrain of the current universe expansion.  
Finally, global 21-cm absorption signal at Cosmic Dawn from the Experiment toDetect the Global EoR Signature (EDGES) opens up a new arena wherein to test cosmographic series \cite{Bowman:2018yin,Yang:2019nhz}.

\section*{Acknowledgements} 

MB and SC acknowledge Istituto Nazionale di Fisica Nucleare (INFN), sezione di Napoli, iniziative specifiche QGSKY and MOONLIGHT-2. 
We also acknowledge the authors of the CosmoMC (A. Lewis) code and Sunny Vagnozzi for the CosmoMC patches on Cosmic Clocks data. This work was developed thanks to the High Performance Computing Center at the Universidade Federal do Rio Grande do Norte (NPAD/UFRN) and the National Observatory (ON) computational support. This paper is based upon work from COST action CA15117 (CANTATA), supported by COST (European Cooperation in Science and Technology).
%%%%%%%%%%%%%%%%%%%%%%%%%%%%%%%%%%%%%%%%%%%%%%%%%%%%%%% 

\begin{table}[h!]
\setlength{\tabcolsep}{1.5em}
%\small
\centering
\caption{$H(z)$ values for different redshift values. $H(z)$ is given by km/s/Mpc.}
\scalebox{1.3}{
{\begin{tabular}{c c c }
\hline
\hline
 $z$ &$H \pm \sigma_H$ &  Ref. \\
\hline
0.0708	& $69.00 \pm 19.68$ & \cite{Zhang14} \\
0.09	& $69.0 \pm 12.0$ & \cite{Jimenez02} \\
0.12	& $68.6 \pm 26.2$ & \cite{Zhang14} \\
0.17	& $83.0 \pm 8.0$ & \cite{Simon05} \\
0.179 & $75.0 \pm	4.0$ & \cite{Moresco12} \\
0.199 & $75.0	\pm 5.0$ & \cite{Moresco12} \\
0.20 &$72.9 \pm 29.6$ & \cite{Zhang14} \\
0.27	& $77.0 \pm 14.0$ & \cite{Simon05} \\
0.28	& $88.8 \pm 36.6$ & \cite{Zhang14} \\
0.35	& $82.1 \pm 4.85$ & \cite{Chuang12}\\
0.352 & $83.0	\pm 14.0$ & \cite{Moresco16} \\
0.3802	& $83.0 \pm 13.5$ & \cite{Moresco16}\\
0.4 & $95.0	\pm 17.0$ & \cite{Simon05} \\
0.4004	& $77.0 \pm 10.2$ & \cite{Moresco16} \\
0.4247	& $87.1 \pm 11.2$  & \cite{Moresco16} \\
0.4497 &	$92.8 \pm 12.9$ & \cite{Moresco16}\\
0.4783	 & $80.9 \pm 9.0$ & \cite{Moresco16} \\
0.48	& $97.0 \pm 62.0$ & \cite{Stern10} \\
0.593 & $104.0 \pm 13.0$ & \cite{Moresco12} \\
0.68	& $92.0 \pm 8.0$ & \cite{Moresco12} \\
0.781 & $105.0 \pm 12.0$ & \cite{Moresco12} \\
0.875 & $125.0 \pm 17.0 $ & \cite{Moresco12} \\
0.88	& $90.0 \pm 40.0$ & \cite{Stern10} \\
0.9 & $117.0 \pm 23.0$ & \cite{Simon05} \\
1.037 & $154.0 \pm 20.0$ & \cite{Moresco12} \\
1.3 & $168.0 \pm 17.0$ & \cite{Simon05} \\
1.363 & $160.0 \pm 33.6$ & \cite{Moresco15} \\
1.43	& $177.0 \pm18.0$ & \cite{Simon05} \\
1.53	& $140.0	\pm 14.0$ & \cite{Simon05} \\
1.75	 & $202.0 \pm 40.0$ & \cite{Simon05} \\
1.965& $186.5 \pm 50.4$ & \cite{Moresco15} \\
\hline
\hline
\end{tabular}}
 \label{tab:Hz}
 }
\end{table} 
%%%%%%%%%%%%%%%%%%%%%%%%%%%%%%%%%%%%%%%%%%%%%%%%%%%%%%%%

\end{document}